\documentclass[%
twocolumn,
 reprint,
 amsmath,amssymb,
 aps,
]{revtex4-2}

\usepackage{graphicx}
\usepackage{dcolumn}
\usepackage{bm}
\usepackage{hyperref}

\hypersetup{
	colorlinks=true,
	linkcolor=blue,
	filecolor=magenta,      
	urlcolor=blue,
	citecolor=blue,
	anchorcolor=blue,
}

\begin{document}

\preprint{APS/123-QED}

\title{Enigmatic 2+6/13 Filling Factor: A Prototype Intermittent Topological State }
\author{Sudipto Das}
\author{Sahana Das}
\author{Sudhansu S. Mandal}
\affiliation{%
	Department of Physics, Indian Institute of Technology, Kharagpur, West Bengal 721302, India
}%

%
%

\date{\today}

\begin{abstract}
	Observation of filling factor 6/13 is one of the surprising fractional quantum Hall states in the second Landau level because, in contrast to the standard wisdom, the fractions ($\nu < 1/2$) with lower numerators, namely 4 and 5, have not yet been observed.  We find that a state indeed forms at $\nu=6/13$ as an intermittent topological state between two prominent states at $\nu =1/2$ and $\nu = 2/5$ with lower numerators. Also, we predict that a  state forms at $\nu=5/13$ as an intermittent to $\nu = 2/5$ and $\nu =3/8$. Our proposed wave functions for $\nu =6/13$ and $5/13$ have excellent overlaps with the corresponding exact ground state wave functions. The Chern-Simons coupling matrices deduced from the form of these wave functions are analyzed to predict the topological properties, which may be experimentally verified.

\end{abstract}


\maketitle

%

The discovery of fractional quantum Hall effect (FQHE) at the even denominator \cite{willett_1987_ObservationEvendenominatorQuantum} filling factor $\nu =1/2$ in the second Landau level (SLL) was astonishing as it did not have any analogue in the lowest Landau level \cite{halperin_1993_TheoryHalffilledLandau}.
This brings an unusual description of the filling factor in terms of the chiral p-wave pairing mechanism \cite{moore_1991_NonabelionsFractionalQuantum} of the composite fermions \cite{jain_1989_CompositefermionApproachFractional,jain_2007_CompositeFermions}, leading to the exotic possibilities such as fractionally charged non-Abelian quasiparticles and an edge mode of charge neutral Majorana fermions. While a general consensus about the true topological order of $\nu =1/2$ state has still not been achieved \cite{rezayi_2011_BreakingParticleHoleSymmetry,	zaletel_2015_InfiniteDensityMatrix,	pakrouski_2015_PhaseDiagramFractional,	rezayi_2017_LandauLevelMixing,	simon_2020_EnergeticsPfaffianAntiPfaffian,
	simon_2018_InterpretationThermalConductance,
	ma_2019_PartialEquilibrationInteger,simon_2020_PartialEquilibrationAntiPfaffian,asasi_2020_PartialEquilibrationAntiPfaffian,
	mross_2018_TheoryDisorderInducedHalfInteger,zhu_2019_DisorderDrivenTransitionFractional,zhu_2020_TopologicalInterfacePfaffian,
	lian_2018_TheoryDisorderedQuantum,mross_2018_TheoryDisorderInducedHalfInteger,wang_2018_TopologicalOrderDisorder,fulga_2020_TemperatureEnhancementThermal}
because of other competitive predicted topological orders \cite{lee_2007_ParticleHoleSymmetryQuantum,levin_2007_ParticleHoleSymmetryPfaffian,son_2015_CompositeFermionDirac,zucker_2016_StabilizationParticleHolePfaffian}, it is believed that $\nu =1/2$ will be the putative state for realizing the non-Abelian quasiparticles and the Majorana mode. Attempts have already been made for realizing such quasiparticles in interferometry, thermal Hall conductivity, and shot noise experiments \cite{lin_2014_RecentExperimentalProgress,banerjee_2018_ObservationHalfintegerThermal,dutta_2022_DistinguishingNonabelianTopological,paul__TopologicalThermalHalla}.
Das {\it et al.} \cite{das_2023_AnomalousReentrantQuantum} have recently proposed an  wave function for describing $\nu =1/2$ FQHE state in the so-called anomalous phase ($\mathcal{A}$ phase).
This wave function predicts \cite{das_2023_AnomalousReentrantQuantum} a downstream charge mode, an upstream neutral mode, and a downstream Majorana mode, subjected to the edge reconstruction. While these edge modes may be equivalent to the modes deduced from particle-hole symmetric Pfaffian wave function \cite{son_2015_CompositeFermionDirac,zucker_2016_StabilizationParticleHolePfaffian}, these two wave functions describe topologically distinct phases. Therefore, the interpretation \cite{banerjee_2018_ObservationHalfintegerThermal,dutta_2022_DistinguishingNonabelianTopological} of the observed thermal Hall conductance remains ambiguous. However, the topological order described by the former wave function \cite{das_2023_AnomalousReentrantQuantum} stands out as the later wave function \cite{zucker_2016_StabilizationParticleHolePfaffian} seems to describe \cite{mishmash_2018_NumericalExplorationTrial} a Fermi liquid \cite{halperin_1993_TheoryHalffilledLandau} of composite fermions.

Several proposals with different sequences \cite{read_1999_PairedQuantumHall,jolicoeur_2007_NonAbelianStatesNegative,bonderson_2008_FractionalQuantumHall,jain_1989_CompositefermionApproachFractional,balram_2018_PartonConstructionWave,balram_2019_PartonConstructionParticleholeconjugate} of states have been made for describing other FQHE states in the SLL as a generalization to Moore and Read Pfaffian \cite{moore_1991_NonabelionsFractionalQuantum} wave function for $\nu=1/2$. However, none of these sequences can describe recently observed \cite{kumar_2010_NonconventionalOddDenominatorFractional,shingla_2018_FinitetemperatureBehaviorSecond} $\nu = 6/13$ FQHE state. The observation of $\nu = 6/13$ state is puzzling as no state ($\nu<1/2$) has been observed with lower numerators, {\it viz.}, 4 and 5. The parton construction which is a generalized \cite{jain_1989_CompositefermionApproachFractional} form of the composite fermion construction for each parton of an electron provides \cite{balram_2018_FractionalQuantumHall} a sequence $\nu = 2n/(5n-2)$ which accommodates $\nu =6/13$ for $n=3$. However, it is not clear why a hole-like $\nu = 2/3$ state features together with some of the particle-like states in a single sequence \cite{balram_2018_FractionalQuantumHall}. There is no such precedence in the lowest Landau level.  
Moreover, moderately high overlap with the proposed wave function for $\nu=6/13$ in the parton model with the exact ground state has been tested \cite{balram_2018_FractionalQuantumHall} for interacting pseudopotentials with zero Landau level mixing (LLM) correction only. On the contrary, the system for which $\nu=6/13$ was observed had LLM parameter $\kappa \simeq 1.2$ \cite{bishara_2009_EffectLandauLevel,peterson_2013_MoreRealisticHamiltonians,sodemann_2013_LandauLevelMixing,simon_2013_LandauLevelMixing}, where $\kappa$ is the ratio between Coulomb and cyclotron energy scales. Consideration of moderate $\kappa$ is essential for describing FQHE states in the SLL as it has recently been shown that topological phase transitions occur \cite{das_2023_AnomalousReentrantQuantum,das_2024_FractionalQuantumHall} for most of the states in the SLL to the $\mathcal{A}$ phase when $ \kappa \gtrsim 1$.

The proposed \cite{das_2023_AnomalousReentrantQuantum} wave function for $\nu =1/2$ has been interpreted as two independent condensates of an equal number of noninteracting composite bosons (CBs) followed by repulsion between CBs of two different condensates such that they feel two zeros at each other's positions. Despite the repulsion between two groups of the CBs, there is no phase separation involved, because symmetrization in particle indices helps to form a liquid state \cite{das_2024_Reply}.
Subsequently, this wave function has been generalized \cite{das_2024_FractionalQuantumHall} to construct wave functions for a sequence of states $\nu = n/(nm-1)$ (for $n\geqslant 1$ and $m\geqslant 3$) in the $\mathcal{A}$ phase.
The sequence \cite{das_2024_FractionalQuantumHall} $\nu = n/(nm-1)$ with their particle-hole conjugate filling factors exhaust all the observed states \cite{shingla_2018_FinitetemperatureBehaviorSecond}, barring $\nu = 6/13$. In this paper, we show that the FQHE state indeed forms at $\nu = 6/13$ as an intermittent state between first two successive states in the hierarchy of $\nu = n/(3n-1)$.

We construct a wave function for the filling factor $\nu =6/13$,  which is in between the first two members in the primary sequence $\nu = n/(3n-1)$ in the second Landau level.
Following a similar construction procedure, we find the formation of an FQHE state also at $\nu = 5/13$ as an intermittent state between the second and third members of the same sequence.
Exact diagonalization of the effective Hamiltonian, including the  LLM corrections \cite{peterson_2013_MoreRealisticHamiltonians} show the presence of $\mathcal{A}$ phase for $\nu =6/13$ and $5/13$ as well. The exact ground state wave functions for these two states in the $\mathcal{A}$ phase have excellent overlaps with the corresponding constructed wave functions. The Chern-Simons coupling matrices \cite{wen_1992_ClassificationAbelianQuantum} deduced from the forms of the constructed wave functions are analyzed to predict the quasiparticle charges and the edge modes.

%
A distinct set of topological orders, which do not have an adiabatic continuation to the topological orders predicted before for the pure Coulomb potentials, are recently predicted \cite{das_2023_AnomalousReentrantQuantum,das_2024_FractionalQuantumHall} for the SLL in the $\mathcal{A}$ phase which occurs in a moderate range of LLM strength $\kappa = (e^2/\epsilon \ell_0)/(\hbar \omega_c)$ as the ratio between the Coulomb and cyclotron energy scales. These FQHE states are formed primarily in the sequence \cite{das_2024_FractionalQuantumHall} of filling factor,
\begin{equation}\label{eq.fillsllgen}
	\nu=n/(nm-1), 
\end{equation}
for $n \geqslant 1$ and $ m \geqslant 3$. The ground state wave functions for these filling factors are well-described by the trial wave functions \cite{das_2024_FractionalQuantumHall}, dropping the ubiquitous Gaussian factor \cite{jain_2007_CompositeFermions,laughlin_1983_AnomalousQuantumHall} in the disk-geometry \cite{haldane_1983_FractionalQuantizationHall},
%
\begin{eqnarray}\label{eq.gen_wavefun}
	&&	\Psi_{\mathcal{A}}\,(n,m)  = \prod_{i<j}^N (z_i-z_j)\, {\cal S} \left[ \prod_{1\leqslant k,l}^{N/(2n)}  \right. \nonumber \\
	&& \left( \prod_{\alpha =0}^{n-1} \left(z_{k+\alpha N/(2n)} -z_{l+N/2+\alpha N/(2n)}\right)^{2(m-2)} \right) \nonumber \\
	& &\left.  \left( \prod_{\alpha\neq \beta,0}^{n-1}
	\left(z_{k+\alpha N/(2n)} -z_{l+N/2+\beta N/(2n)}\right)^{2(m-1)}  \right) \right] \, .
\end{eqnarray}
Here, $ z_j = (x_j - i y_j)/\ell_0 $ is the complex particle coordinates with $ \ell_0 $ being the magnetic length and $ \mathcal{S} $ represents symmetrization with respect to all $ N $ particle indices. The wave function in Eq.~\eqref{eq.gen_wavefun} corresponds to the the total angular momentum $ M = N (\nu^{-1}N - 1)/2 $ for an $N$ particle system.
The Jastrow factor $ J=\prod_{i<j}^N (z_i-z_j) $ in Eq.~\eqref{eq.gen_wavefun} describes the attachment of one flux quantum to each electron, converting it into a composite particle \cite{jain_2007_CompositeFermions} as a bound-state of an electron with one unit of flux quantum. An interchange between two such particles provides an additional phase of $\pi$ and hence these particles are the CBs. 
This wave function is interpreted \cite{das_2024_FractionalQuantumHall} as the formation of two groups of CBs with strong mutual repulsion. Since the CBs find options in joining any of the two groups, the wave function has a hidden \cite{cappelli_2001_ParafermionHallStates,das_2023_AnomalousReentrantQuantum} $\mathbb{Z}_2$ symmetry.  
For the given series in Eq.~\eqref{eq.fillsllgen} each condensate is now subdivided into $n$ fictitious sectors with an equal number of CBs and each CB of a sector feels $2(m-2)$ zeros at the positions of other CBs of a sector and $2(m-1)$ zeros at the position of CBs at all other $(n-1)$ sector in the other group.
Indistinguishability of the CBs which is manifested by explicit symmetrization of the arrangement of the CBs allows them to form a liquid phase because two groups can be formed with any $N/2$ of those particles.

Amongst the observed filling factors \cite{kumar_2010_NonconventionalOddDenominatorFractional} in the SLL, only $\nu = 6/13$ does not belong to the sequence of filling factors \eqref{eq.fillsllgen} or their particle-hole conjugate partners. Recall that the filling factors like $\nu = 4/11$ and $5/13$ do not belong to the Jain sequences \cite{jain_1989_CompositefermionApproachFractional,jain_2007_CompositeFermions} of filling factors in the lowest Landau level. These states are in between two prominent states 1/3 and 2/5 in the principal Jain sequence, and are understood \cite{mukherjee_2014_Enigmatic11State} as FQHE of the composite fermions in their partially filled effective Landau level with an unusual mechanism. We here show the formation of intermittent states such as $\nu =6/13$ and $5/13$ between two successive prominent states in the sequence of filling factors \eqref{eq.fillsllgen}.

For every $m$ in Eq.~\eqref{eq.fillsllgen}, the first two filling factors in the sequence are $1/(m-1)$ and $2/(2m-1)$. The wave functions in Eq.~\eqref{eq.gen_wavefun} for $\nu = 1/(m-1)$ are interpreted in terms of two condensates, each having $N/2$ CBs. For $\nu = 2/(2m-1)$, each condensate is further divided into two sectors, each of which contains $N/4$ CBs. We now construct a state (for $m=3$) as an intermittent to $1/2$ and $2/5$ where two sectors in a condensate will have an unequal number of CBs, in the ratio of $1:2$, {\em i.e.}, in each condensate, one sector contains $N/3$ and the other sector contains $N/6$ CBs. Further, CBs in two sectors with $N/3$ CBs in two different condensates mutually feel $2$ zeros at each other's positions. Similarly, CBs in two sectors with $N/6$ CBs mutually feel $1$ zero. Finally, CBs in a sector of $N/3$ CBs mutually feel $3$ zeros at the positions of CBs in the sector of $N/6$ CBs in the other condensate, and vice versa. This description leads us to write a wave function 
\begin{eqnarray}\label{eq.wave613}
&&		\Psi_{\mathcal{A}}^{\frac{6}{13}}  = J \times
	{\cal S} \left[ \left(\prod_{1\leqslant k,l}^{N/3} \left( z_k - z_{l+N/2} \right)^{2}\right) \right.  \nonumber \\
	&&		 \times    \left( \prod_{1\leqslant k }^{N/3} \, \prod_{1\leqslant l}^{N/6} \prod_{\gamma \neq \bar{\gamma}, 0 }^{1} \left( z_{k + \gamma N/2} - z_{l+ \bar{\gamma} N/2+N/3} \right)^3 \right)  \nonumber \\
	&&		\left. \times	\left( \prod_{1\leqslant k,l}^{N/6} \left( z_{k+N/3} - z_{l+N/2+N/3} \right) \right) \right]\, , 
\end{eqnarray}
(see Ref.~\onlinecite{Suppli} for its schematic representation) which turns out to be a wave function for the filling factor $\nu = 6/13$.
The total angular momentum of the wavefunction \eqref{eq.wave613} for $\nu = 6/13$ is thus $M = N(13N/6-1)/2$. Therefore, the relationship between the number of flux quanta and the number of electrons for $\nu=6/13$ is $N_\Phi = 13N/6 -1$ with the flux-shift $S=1$ in a spherical geometry. In fact, all the wave functions in Eq.\eqref{eq.gen_wavefun} correspond to $S=1$. The construction of the wave function ensures that it will have zero total angular momentum ($L=0$) in spherical geometry.

Additionally, we construct a wave function for $\nu = 5/13$, which forms as an intermittent state between two observed states $\nu =2/5$ and $3/8$ in the sequence $\nu = n/(3n-1)$ in Eq.~\eqref{eq.fillsllgen}. It is given by \cite{Suppli}
\begin{eqnarray}\label{eq.wave513}
		\Psi_{\mathcal{A}}^{\frac{5}{13}}  = J
	 \times  {\cal S} \left[ \left( \prod_{1\leqslant k,l}^{N/5}  \right.
		\prod_{\alpha =0}^{1} \left(z_{k+\alpha N/5} -z_{l+N/2+\alpha N/5}\right)^{2}  \right) &&
		\nonumber \\
			\times \left( \prod_{1\leqslant k}^{N/10} \prod_{1\leqslant l}^{N/5} \prod_{\alpha = 0}^{1} \prod_{\gamma \neq \bar{\gamma}, 0}^{1} \left(z_{k+ \gamma N/2 + 2N/5} -z_{l+ \bar{\gamma} N/2+\alpha N/5}\right)^{4} \right) &&  \nonumber \\
			\times  \left. \left(\prod_{1\leqslant k,l}^{N/5}  \prod_{\alpha \neq\beta,0}^{1}
		\left(z_{k+\alpha N/5} -z_{l+N/2+\beta N/5}\right)^{4} \right)  \right]. && 
\end{eqnarray}
There are three sectors having $N/5,\,N/5$, and $N/10$ CBs, respectively, in each of the two condensates, as described before. A sector with $N/5$ CBs in a condensate mutually feels 2 zeros at each other's position of one sector of $N/5$ CBs and 4 zeros at each other's position of the remaining CBs of two sectors in the other condensate. However, the CBs in the sectors with $N/10$ in both condensates do not feel any repulsion between them. In spherical geometry, the wave function $\Psi_{\mathcal{A}}^{\frac{5}{13}}$ also corresponds to $S=1$, {\em i.e.,} $N_\Phi = 13N/5 -1$.

We now numerically show the existence of FQHE states $\nu =6/13$ and $\nu=5/13$ corresponding to $S=1$ for spherical geometry in the $\mathcal{A}$ phase, followed by the accuracy of the constructed wave functions \eqref{eq.wave613} and \eqref{eq.wave513}.

\begin{figure}
	\centering
	\includegraphics[width=\linewidth]{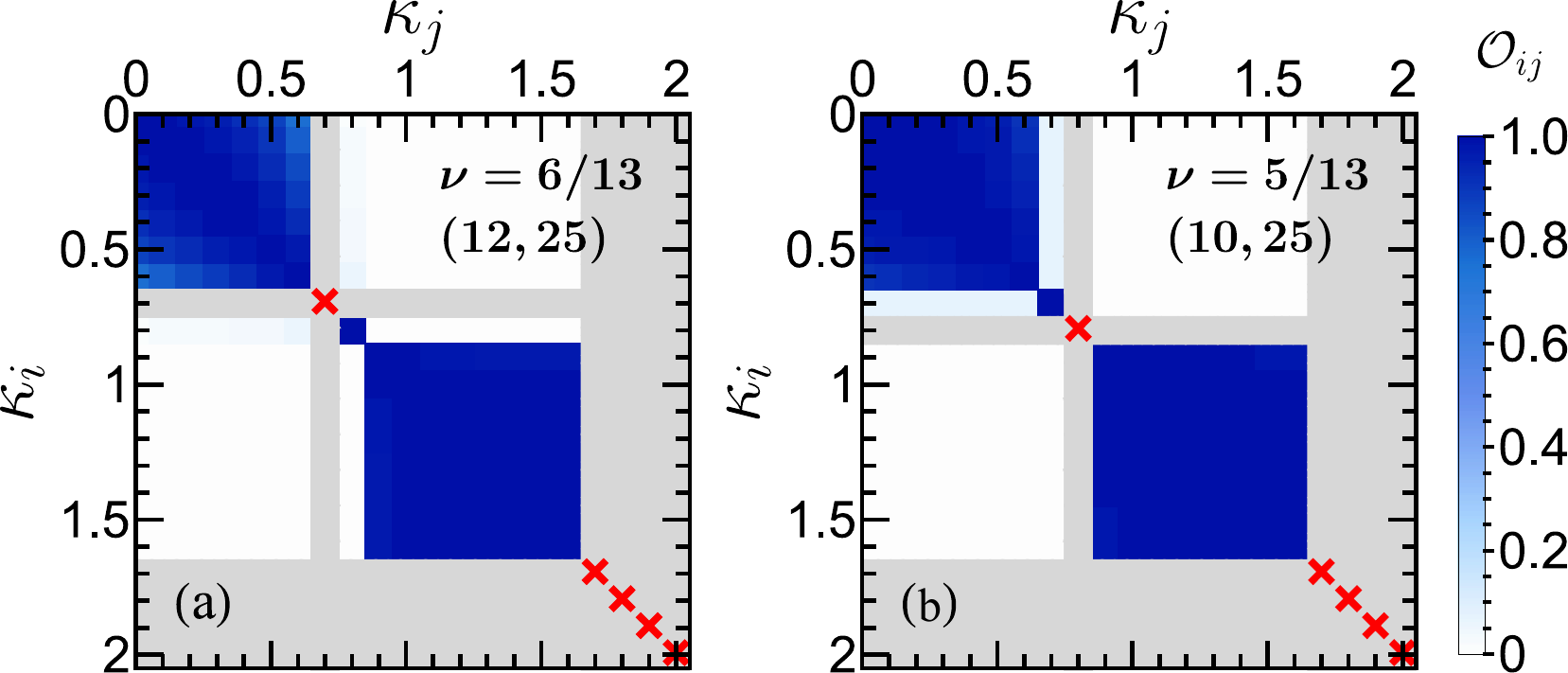}
	\caption{(color online) (a) The color map indicates the overlaps of the exact ground states (if found at $L=0$) of the Hamiltonian in the Eq.~\eqref{eq.hamiltonianLLM} with LLM-corrected pseudopotentials \cite{pseudopotential} at the filling factor 6/13 for $N=12$ particles at flux, $N_\Phi = 25$. If a ground state is not found at $L=0$ the state is un-quantized and marked with a red cross, and the overlap regime of that state with all other ground states at different $\kappa$ is grayed out.
		(b) Same as (a) but for the filling factor 5/13 with $N=10$ particles at flux, $N_\Phi = 25$.}
	\label{fig.phaseplotsll613}
\end{figure}

The effective Hamiltonian for the fully spin-polarized electrons in the SLL  incorporating the effects of LLM can be written \cite{bishara_2009_EffectLandauLevel,peterson_2013_MoreRealisticHamiltonians} in spherical geometry \cite{haldane_1983_FractionalQuantizationHall} as,
\begin{eqnarray}\label{eq.hamiltonianLLM}
	\hat{H}_{\text{eff}}(\kappa) &=&
	\sum_{\lambda\, \text{odd}} \left[ V_\lambda^{(2)} + \kappa \,\delta V_\lambda^{(2)} \right] \sum_{i<j} \hat{P}_{ij}(\lambda) \nonumber \\
	&&	+ \sum_{\lambda\geqslant 3}
	\kappa \, V_\lambda^{(3)} \sum_{i<j<k} \hat{P}_{ijk}(\lambda) \, ,
\end{eqnarray}
where $ V_\lambda^{(2)} $ represents two-body Coulomb pseudo-potential in the SLL; $ \delta V_\lambda^{(2)} $ and $ V_\lambda^{(3)} $ are respectively the two-body pseudo-potential corrections and three-body pseudo-potentials arising due to the LLM whose strength is determined by $\kappa$.
Here $ \hat{P}_{ij}(\lambda) $ and $ \hat{P}_{ijk}(\lambda) $ are two and three-body projection operators, respectively, onto pairs and triplets of electrons with relative angular momentum $ \lambda $.
We perform exact diagonalization of the effective Hamiltonian in Eq.~\eqref{eq.hamiltonianLLM} for different $ \kappa $ in the spherical geometry with the LLM corrected pseudopotentials \cite{pseudopotential} for the filling fractions $\nu =6/13$ and 5/13 for a finite system of $ N $ particles with flux-quanta, $ N_\Phi = \nu^{-1}N - 1 $.
For different values of $\kappa$, if the exact ground state is found at the total angular momentum $L=0$, we determine the overlap of those exact ground states as $\mathcal{O}_{ij} = \langle \Psi_{\text{gs}}(\kappa_i) \mid \Psi_{\text{gs}}(\kappa_j) \rangle $.
In Fig.~\ref{fig.phaseplotsll613}, two topologically distinct phases can be identified at different ranges of $\kappa$ as the overlap between the exact ground states belonging to two distinct phases is zero, implying the phases are mutually orthogonal to each other. The overlaps of the exact ground states at two different $\kappa$ belonging to the same topological phase is nearly unity.
Thus, alike \cite{das_2023_AnomalousReentrantQuantum,das_2024_FractionalQuantumHall} all other states in the SLL, the reentrant $\mathcal{A}$ phase at the moderate range of $\kappa \sim 0.9-1.6 $ persists at these two unconventional filling fractions as well.

\begin{figure}
	\centering
	\includegraphics[width=\linewidth]{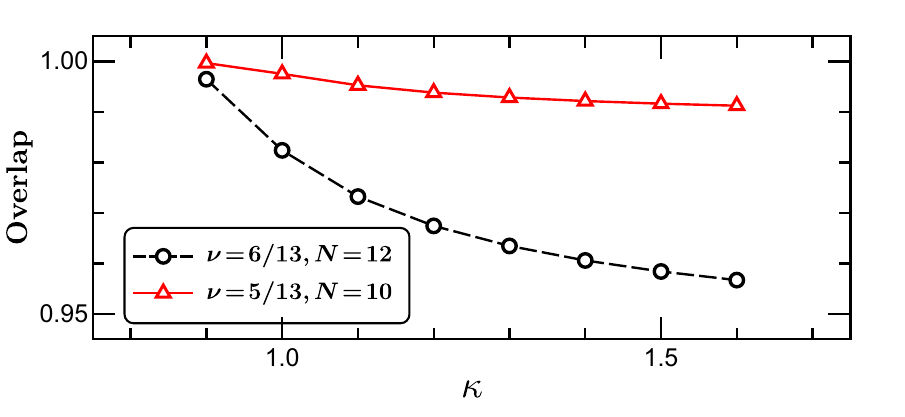}
	\caption{(color online) Overlaps of the proposed wave functions \eqref{eq.wave613} and \eqref{eq.wave513} for  6/13 and 5/13 states respectively with their respective exact ground states in the $\mathcal{A}$ phase. }
	\label{fig.ovlpplotsll613}
\end{figure}

In Fig.~\ref{fig.ovlpplotsll613}, we show very high overlap of our proposed trial wave functions in Eqs.~\eqref{eq.wave613} and \eqref{eq.wave513} at filling factors 6/13 and 5/13, respectively, with the corresponding exact ground states in the $\mathcal{A}$ phase. We, therefore, believe that the proposed wave functions capture the topological order of the ground states of the corresponding filling factors. As these wave functions do not vanish even when $N/2$ CBs acquire the same position \cite{read_1999_PairedQuantumHall,cappelli_2001_ParafermionHallStates}, we believe that the ground states in the $\mathcal{A}$ phase for these two filling factors will support non-Abelian quasiparticles as well. Recall that non-Abelian MR wave function has a similar property that two CBs can occupy the same position when one factors out a Jastrow factor, which is the minimal requirement to respect the Pauli exclusion principle for electrons.

\begin{figure}
	\centering
	\includegraphics[width=\linewidth]{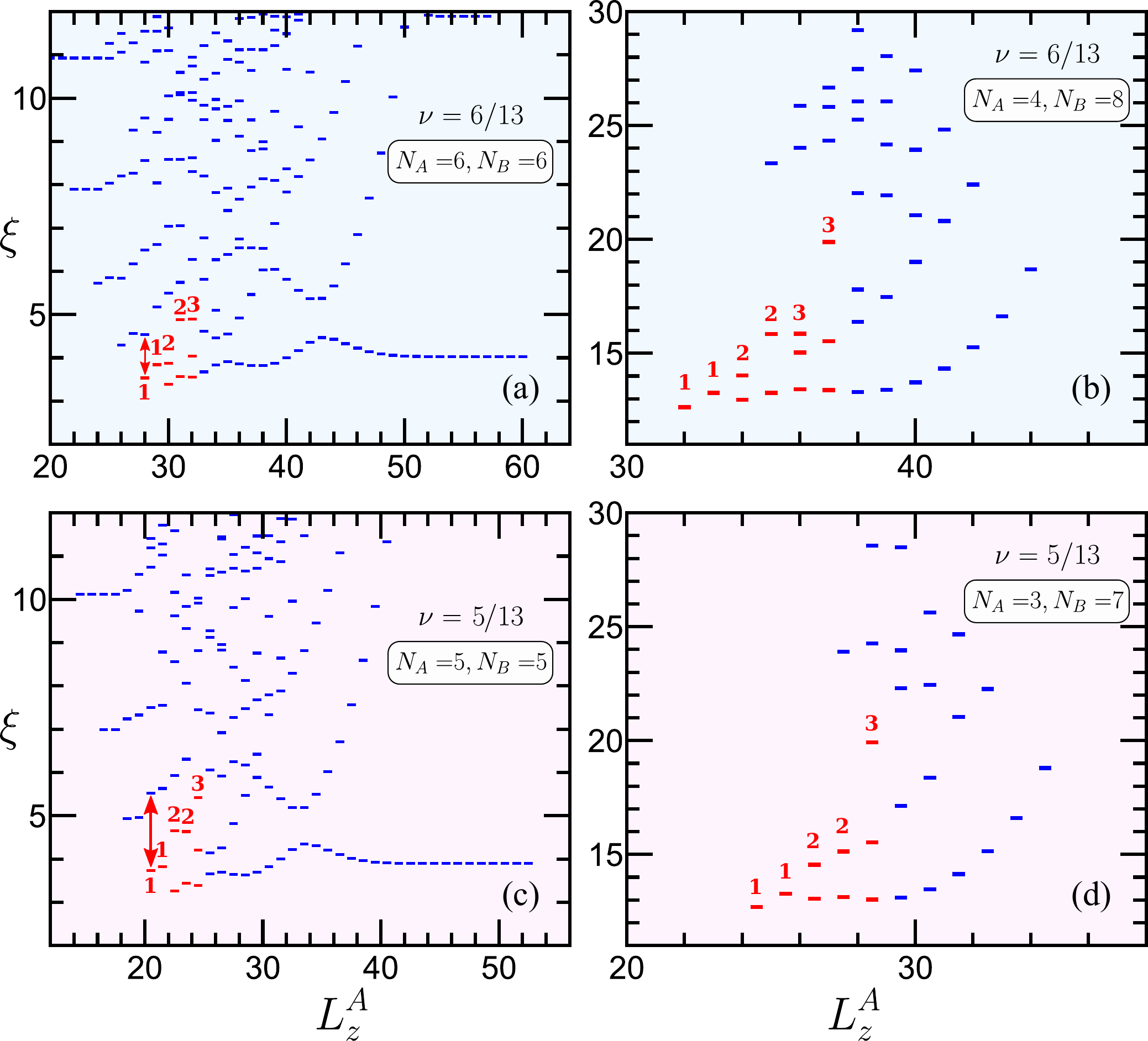}
	\caption{(color online) Entanglement Spectrum for $ \nu= $ 6/13 and 5/13 states in the $ \mathcal{A} $ phase ($ \kappa =1.2 $) for an equal and an unequal number of particles $ N_A $ and $ N_B $ in two partitions $A$ and $B$, {\it i.e.,} two hemispheres. $L_Z^A$ is the sum of the azimuthal components of angular momentum occupied by the $N_A$ particles of $A$ part of the partition, and $\xi$ represents the entanglement energy in an arbitrary unit.}
	\label{fig.esf613f513}
\end{figure}

In Fig.~\ref{fig.esf613f513}, we show the entanglement spectra (ES) for the exact ground states at the $\mathcal{A}$ phase for the filling factors $\nu = 6/13$ and $5/13$. 
The ES has been obtained by orbital partition  \cite{das_2023_AnomalousReentrantQuantum} at the equator of the sphere on which the electrons reside for both equal and unequal numbers of particles in two partitions.
The sequence of the counting of low-lying states 1-1-2-2-3-3 is not only the same for both $\nu=6/13$ and 5/13, but this sequence is also universal \cite{das_2024_FractionalQuantumHall} to all the FQHE states in the $\mathcal{A}$ phase.

\begin{figure}[t]
	\centering
	\vspace{0.5cm}
	\includegraphics[width=1\linewidth]{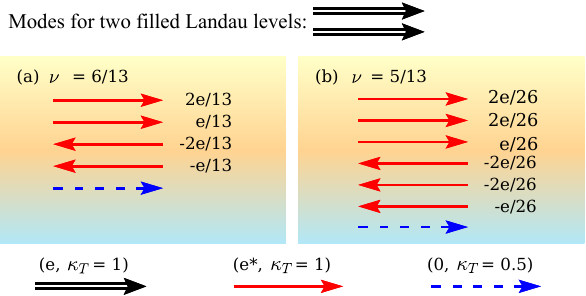}
	\caption{Schematic of edge modes of 6/13 and 5/13 states in the SLL.
		The charge modes (represented as different arrow lines) as obtained from the $\mathbb{K}$-matrices in Eq.~\eqref{eq.Kmat}.
		Two arrows on the top are due to two filled Landau levels.
		Vivid edge modes represented by the arrows, labeled at the bottom, follow as, solid-double line: electronic mode, solid-single line: charged quasiparticle mode,  dashed line: Majorana mode. The corresponding quasiparticle/quasihole charge, $e^*$, and thermal Hall conductivity, $\kappa_T$ are mentioned above the corresponding arrows.
	}
	\label{fig.edgemodessll613}
\end{figure}

Study of low energy effective Chern-Simons Lagrangian density \cite{wen_1992_ClassificationAbelianQuantum},
\begin{equation}\label{eq.lagrangian}
	\mathcal{L} = -\frac{1}{4\pi}\epsilon^{\alpha\beta\gamma} \sum_{I,J=1}^{M} \mathbb{K}_{IJ} a_\alpha^I \partial_\beta a_\gamma^J - \frac{1}{2\pi}\epsilon^{\alpha\beta\gamma} \sum_{I=1}^{K} t_I A_\alpha \partial_\beta a_\gamma^I
\end{equation}
can serve as a great tool to unveil the topological properties of the $ \mathcal{A} $ phase of 6/13 and 5/13 states. Here, $a_\alpha^I$ represents $I$-th component of $M=4$- and 6-component Chern-Simons gauge fields reminiscent of the total sectors for the wave functions in Eqs.\eqref{eq.wave613} and \eqref{eq.wave513} respectively. $A_\alpha$ is the external electromagnetic field, and $\epsilon^{\alpha\beta\gamma}$ is the antisymmetric Levi-Cevita tensor. The proposed wave functions in Eqs.~\eqref{eq.wave613} and \eqref{eq.wave513} for 6/13 and 5/13 states, respectively,  provide the symmetric $\mathbb{K}$-matrices \cite{Suppli} as,
\begin{equation}\label{eq.Kmat}
	\mathbb{K}_{6/13} = 	\begin{pmatrix}
		1 & 1 & 3 & 4 \\
		1 & 1 & 4 & 2 \\
		3 & 4 & 1 & 1 \\
		4 & 2 & 1 & 1 \\
	\end{pmatrix} \,,
	\, \mathbb{K}_{5/13} = 	\begin{pmatrix}
		1 & 1 & 1 & 3 & 5 & 5 \\
		1 & 1 & 1 & 5 & 3 & 5 \\
		1 & 1 & 1 & 5 & 5 & 1 \\
		3 & 5 & 5 & 1 & 1 & 1 \\
		5 & 3 & 5 & 1 & 1 & 1 \\
		5 & 5 & 1 & 1 & 1 & 1 \\
	\end{pmatrix}
\end{equation}
Following Ref.~\cite{wen_1992_ClassificationAbelianQuantum} we introduce the charge vectors $t^T = (1,1,1,1)$ and $t^T = (1,1,1,1,1,1)$ for $\nu =t^T \mathbb{K}_{6/13}^{-1} t = 6/13$ and $ \nu =t^T \mathbb{K}_{5/13}^{-1} t =5/13$ respectively. The number of edge modes will be the same as the dimension of $\mathbb{K}$ matrices, {\it i.e.}, there will be $4$ and $6$ edge modes respectively for $\nu = 6/13$ and 5/13 carrying respective total charges $6e/13$ and $5e/13$. As there are an equal number of positive and negative eigenvalues for both $\mathbb{K}_{6/13}$ and $\mathbb{K}_{5/13}$, there will be 2  for $\nu=6/13$
and 3 for $\nu=5/13$ downstream edge modes of quasiparticles and an equal number of upstream edge modes of quasiholes. While the quasiparticle vectors $l^T = (1,0,0,0)$ and $ (0,0,1,0)$ for $\nu = 6/13$ produces same quasiparticle charge $q = el^T \mathbb{K}_{6/13}^{-1} t = 2e/13$, the quasiparticle vectors $l^T = (0,1,0,0)$ and
$l^T = (0,0,0,1)$ provides same and the lowest quasiparticle charge $q= e/13$.
Similarly, for $\nu=5/13$ state, the quasiparticle vectors $l^T=(1,0,0,0,0,0)$, $(0,1,0,0,0,0)$, $(0,0,0,1,0,0)$, and $(0,0,0,0,1,0)$ produce quasiparticle charge $q= 2e/26 $, and the lowest quasiparticle charge of $q= e/26$ is obtained from quasiparticle vectors of $l^T= (0,0,1,0,0,0)$ and $(0,0,0,0,0,1)$. All these edge modes with corresponding quasiparticle charges are depicted in Fig.~\ref{fig.edgemodessll613}. In addition, one neutral Majorana mode will be present for both 6/13 and 5/13 states because of the hidden \cite{cappelli_2001_ParafermionHallStates,das_2023_AnomalousReentrantQuantum} $\mathbb{Z}_2$ symmetry present in the wave functions in Eqs.~\eqref{eq.wave613} and \eqref{eq.wave513}.

%
In summary, the experimentally observed $\nu = 6/13$ in the second Landau level is shown to form in the so-called ${\mathcal A}$ phase. The construction of the proposed wave function, which does possess excellent overlap with the exact ground state, reveals that it so forms as an intermittent state between two first prominent states $\nu =1/2$ and $\nu = 2/5$ in the primary sequence $\nu = n/(3n-1)$. We further predict the $\nu = 5/13$ state as an intermittent state between the observed $\nu = 2/5$ and $\nu = 3/8$ state in the same sequence. Experimental realization of $\nu = 5/13$ state will strengthen our theory, as no theory has thus far predicted this state. Also, we predict the charges of the quasiparticles and the edge modes for both these states for experimental verifications.

%
We thank the developers of DiagHam software, which has been extensively used in this work. We acknowledge the supercomputing facility provided by
Param Shakti (IIT Kharagpur), a National Supercomputing Mission, Government of India.


%

\clearpage

\renewcommand{\figurename}{FIG. S\!\!}
\renewcommand{\tablename}{TABLE S\!}
\renewcommand*{\thesection}{S\arabic{section}}
\renewcommand\theequation{S\arabic{equation}}
\renewcommand{\bibnumfmt}[1]{[S#1]}

\setcounter{equation}{0}
\setcounter{figure}{0}

\onecolumngrid

\begin{widetext}
	\centering
	\Large{\bf{ Supplemental Material for ``Enigmatic 2+6/13 Filling Factor: A Prototype Intermittent Topological State" } }\\[1cm]
\end{widetext}

\begin{figure}[h]
	\centering
	\includegraphics[width=0.9\linewidth]{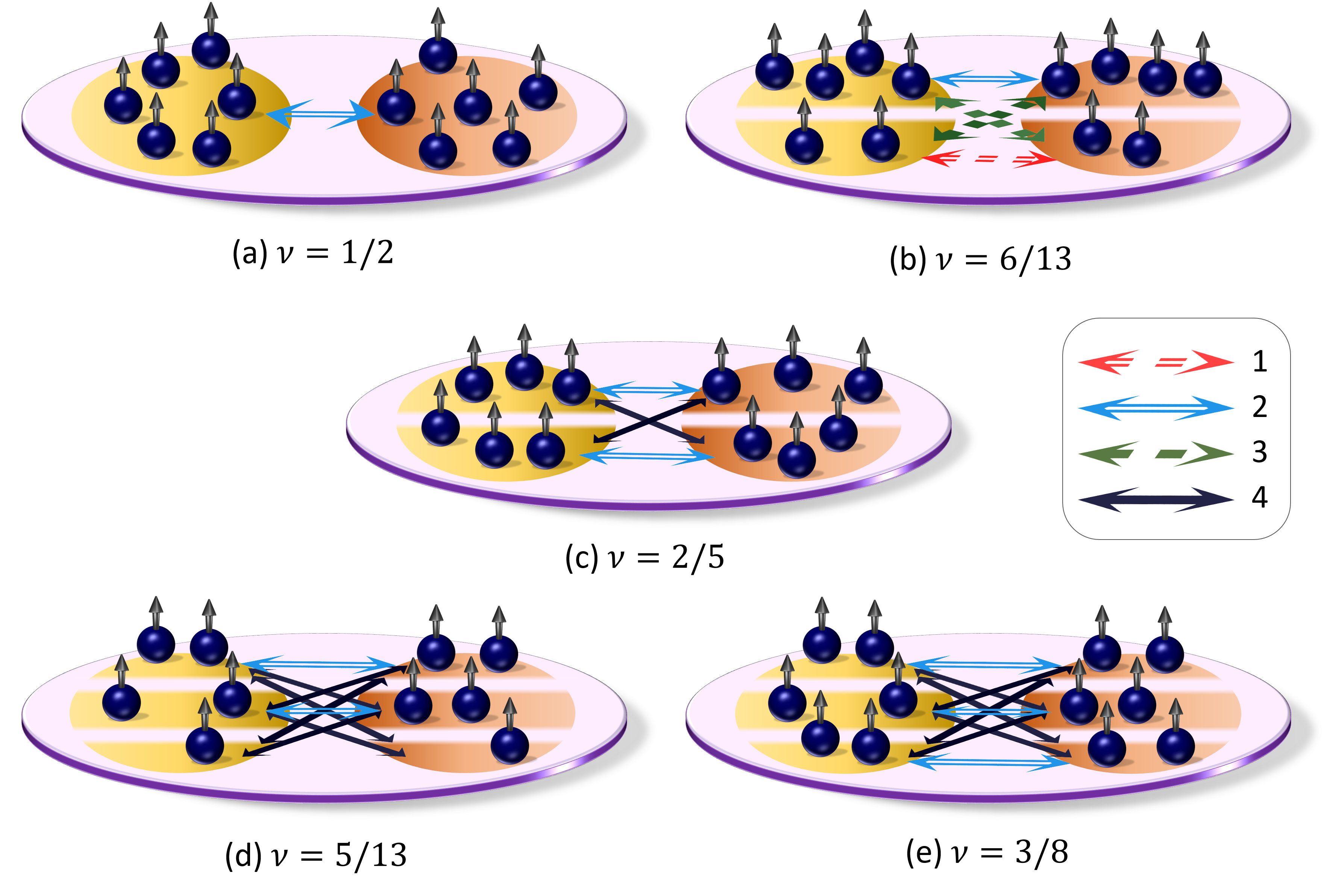}
	\caption{Schematic of the constructed ground state wave functions
		for $\nu= 1/2, 6/13, 2/5, 5/13, 3/8$ states respectively in (a)--(e). 
		The blue balls represent the electrons and the attached arrow to it represents one flux quantum attached to the electrons; as a whole the balls with an arrow represent the composite bosons (CBs).
		A system of $N$ CBs forms two groups of $ N/2 $ each; these two groups are represented by the yellow and orange shaded regions. 
		In (b)--(e), each group is further subdivided into fictitious sectors;  horizontal strips in the shaded regions are shown as dividers between sectors.
		CBs belonging to different sectors of two groups feel strong repulsion, which is shown by the double headed arrow symbols, and the corresponding strength of repulsion is represented by different kinds of arrow-heads: red double-dashed arrow, cyan double arrow, green dashed arrow, and blue arrow are for repulsion strengths  1, 2, 3, and 4 respectively. These strengths are the corresponding exponents in the wave functions.
	}
	\label{fig.sllall}
\end{figure}

\onecolumngrid
The Jastrow factor in  Eq.~(2), $\prod_{i<j}^N (z_i-z_j)$, is the minimal requirement for the Pauli exclusion principle for electrons. However, factoring out this can be thought as an attachment of one unit flux to each electron, transforming them into the CBs. Now, the remaining factor of the wave function (excluding the symmetrization of the indices) can be described as follows: First, they are separated into two groups having an equal number of CBs; Second, each group is further subdivided into $n$ sectors; Third, the CBs belonging to a sector of a group feel repulsion with the CBs of the other group. The corresponding strength of repulsion can be read from the exponents of $(z_k-z_l)$ where $k$ and $l$ indices correspond to CBs of two different groups. Figures~S\ref{fig.sllall}(a), S\ref{fig.sllall}(c), and S\ref{fig.sllall}(e), respectively are the schematic representation of this description for 1/2, 2/5, and 3/8 states. The symmetrization of indices in Eq.~(2) takes care of the indistinguishability of the CBs.

Similarly, the wavefunction (3) for 6/13 can be understood as two groups of CBs; each group is further subdivided into two sectors; the number of CBs in two sectors of a group is in the ratio 1:2; the corresponding strength of repulsion between CBs can be read of from the exponents of $(z_k-z_l)$, as depicted in Fig.~S\ref{fig.sllall}(b). As $1/2 > 6/13 >2/5$ and there are two sectors in each group with equal number of CBs for forming 2/5 state, 6/13 state forms with two sectors but with an unequal number of CBs in a group. Therefore, 6/13 can be referred to as an intermittent state: A transient state that forms as a prematured daughter state of primary sequence. 

Figure~S\ref{fig.sllall}(d) is the schematic depiction of the wavefunction (4) for the 5/13 state. As $2/5 > 5/13 > 3/8$, 5/13 state forms with 3 sectors in each group of CBs. The number of CBs in these sectors is in the ratio 1:2:2, while 3/8 state forms with the ratio 1:1:1.

\subsection*{Construction of $\mathbb{K}$ Matrix:}

The form of $\mathbb{K}$ matrices corresponding to wave functions in Eqs.(2)--(4), can be obtained trivially from the corresponding diagrams in Fig.~S\ref{fig.sllall}. The dimension of $\mathbb{K}$ matrix will be equal to the total number of sectors of two groups of CBs. The elements of a $\mathbb{K}$ matrix are found out as follows: The element $K_{ij}$ can be read as the repulsion strength between CBs of $i^{\rm th}$ and $j^{\rm th}$ sectors; In addition, each of the elements should be increased by 1 for the Jastrow factor that has been factored out for transforming electrons into CBs. Therefore, the $\mathbb{K}$ matrices for states $1/2$, $2/5$ and $3/8$ states are found to be

\begin{equation}\label{eq.k 2/5}
	\mathbb{K}_{1/2}= \begin{pmatrix}
		1 & 3\\
		3 & 1
	\end{pmatrix}
	\quad\text{,}\quad
	\mathbb{K}_{2/5}= \begin{pmatrix}
		1 & 1 & 3 & 5\\
		1 & 1 & 5 & 3\\
		3 & 5 & 1 & 1\\
		5 & 3 & 1 & 1
	\end{pmatrix}
	\quad\text{and}\quad
	\mathbb{K}_{3/8}= \begin{pmatrix}
		1 & 1 & 1 & 3 & 5 & 5\\
		1 & 1 & 1 & 5 & 3 & 5 \\
		1 & 1 & 1 & 5 & 5 & 3 \\
		3 & 5 & 5 & 1 & 1 & 1 \\
		5 & 3 & 5 & 1 & 1 & 1 \\
		5 & 3 & 3 & 1 & 1 & 1 
	\end{pmatrix}
\end{equation}

Similarly, $\mathbb{K}_{6/13}$ and $\mathbb{K}_{5/13}$ in Eq.~(7) can be read out from Figs.~S\ref{fig.sllall}(b) and S\ref{fig.sllall}(d) respectively.

\end{document}